\documentclass{aa}  

\usepackage{graphicx}
\usepackage{txfonts}
\usepackage[version=3]{mhchem}
\usepackage{xcolor}
\usepackage{lscape}
\usepackage{ulem}
\usepackage{mhchem}

\usepackage{booktabs}
\usepackage{hyperref}

\usepackage[figuresright]{rotating}
\usepackage{mathtools}
\usepackage{float}
\usepackage{csquotes}

\begin{document}

   \title{Assessing realistic binding energies of some essential interstellar radicals with amorphous solid water}

   \subtitle{A fully quantum chemical approach}

   \author{M. Sil\inst{\ref{inst1},\ref{inst2},\ref{inst3}} \and A. Roy\inst{\ref{inst4},\ref{inst3}} \and P. Gorai\inst{\ref{inst5},\ref{inst6},\ref{inst3}} \and N. Nakatani\inst{\ref{inst7}} \and T. Shimonishi\inst{\ref{inst8}} \and K. Furuya\inst{\ref{inst9},\ref{inst12}} \and N. Inostroza-Pino\inst{\ref{inst10}} \and P. Caselli\inst{\ref{inst11}} \and A. Das\inst{\ref{inst3},\ref{inst11}}}

   \institute{Univ. Grenoble Alpes, CNRS, IPAG, 38000 Grenoble, France \\
              \email{milansil93@gmail.com}
              \label{inst1}
         \and
             Univ Rennes, CNRS, IPR (Institut de Physique de Rennes) - UMR 6251, F-35000 Rennes, France
             \label{inst2}
         \and
             Department of Chemical Sciences, Indian Institute of Science Education and Research Kolkata, Mohanpur 741246, West Bengal, India \\
             \email{aroy.astro@gmail.com}
             \label{inst4}
         \and
             Rosseland Centre for Solar Physics, University of Oslo, PO Box 1029 Blindern, 0315 Oslo, Norway
             \label{inst5}
         \and
             Institute of Theoretical Astrophysics, University of Oslo, PO Box 1029 Blindern, 0315 Oslo, Norway
             \label{inst6}
        \and
             Institute of Astronomy Space and Earth Sciences, P 177, CIT Road, Scheme 7m, Kolkata 700054, West Bengal, India \\
             \email{ankan.das@gmail.com}
             \label{inst3}     
         \and
            Department of Chemistry, Graduate School of Science and Engineering, Tokyo Metropolitan University, 1-1 Minami-Osawa, Hachioji, Tokyo 192-0397, Japan
            \label{inst7}    
         \and
            Institute of Science and Technology, Niigata University, Ikarashi-nihoncho 8050, Nishi-ku, Niigata 950-2181, Japan
            \label{inst8}
        \and
            National Astronomical Observatory of Japan, Osawa 2-21-1, Mitaka, Tokyo 181-8588, Japan
            \label{inst9}
        \and
            Department of Astronomy, Graduate School of Science, University of Tokyo, Tokyo 113-0033, Japan
            \label{inst12}
        \and
            Universidad Autónoma de Chile, Facultad de Ingeniería, Núcleo de Astroquímica \& Astrofísica, Av. Pedro de Valdivia 425, Providencia, Santiago, Chile
            \label{inst10}
        \and
             Max-Planck-Institute for extraterrestrial Physics, P.O. Box 1312 85741 Garching, Germany
             \label{inst11}}


 
  \abstract 
{In the absence of laboratory data, state-of-the-art quantum chemical approaches can provide estimates of the binding energy (BE) of interstellar species with grains. Without BE values, contemporary astrochemical models are compelled to utilize wild guesses, often delivering misleading information.
Here, we employed a fully quantum chemical approach to estimate the BE of seven diatomic radicals -- CH, NH, OH, SH, CN, NS, and NO -- that play a crucial role in shaping the interstellar chemical composition, using {a suitable} amorphous solid water model as a substrate since water is the principal constituent of interstellar ice in dense and shielded regions.
While {the} BEs are compatible with physisorption, the binding of CH in some sites shows chemisorption, in which a chemical bond to an oxygen atom of a water molecule is formed. While no structural change has been observed for the CN radical, it is believed that the formation of a hemibonded system between the outer layer of the water cluster and the radical is the reason for the unusually large BE in one of the binding sites considered in our study. A significantly lower BE for NO, consistent with recent calculations, is obtained, which helps explain the recently observed HONO/NH$_2$OH and HONO/HNO ratios in the low-mass hot corino IRAS 16293–2422~B with chemical models.
}
   \maketitle
\section{Introduction \label{sec:intro}}
Currently, more than 300 molecules have been detected in the interstellar medium or circumstellar shells\footnote{\url{https://cdms.astro.uni-koeln.de/classic/molecules}} \citep{mcgu22}.
Interstellar grains act as catalysts in shaping the chemical composition of interstellar clouds \citep{hase92,das08a,das10,das16,das19,das21,das11,sriv22,sil18,sil21,ghos22,mond21}.
$\rm{H_2O}$ is the most dominant ice component in dense and cold regions of interstellar clouds \citep{gibb04,boog15}. It can account for $60-70$\% of the ice along most lines of sight \citep{whit03}.
In dense and shielded regions, multiple layers of ice would form either via the direct adsorption of gas-phase species or via chemical processing on the grain surface.
The proper structure of interstellar ice remains unclear, although it is generally agreed that it is amorphous upon formation \citep{hama13}.

Estimating the binding energies (BEs) of interstellar species is essential for interpreting observations and reconstructing our astrochemical history.
Several recent studies attempted to provide theoretical BEs of interstellar species.
\cite{sil17} calculated the BE of H and $\rm{H_2}$ (the most abundant species) on different types of grain surfaces and found {that} the BE of $\rm{H_2}$ is always higher than that of H;
\citet[see their Table~2]{wake17} reported the BEs of $79$~interstellar species, including radicals, using a single $\rm{H_2O}$ molecule (monomer) as an approximate amorphous solid water (ASW) ice surface based on quantum chemical calculations.
They compared the calculated BEs with experimental values for some molecules. They conclude that the calculated values of the physisorbed BE considering a single $\rm{H_2O}$ molecule as substrate are proportional to the experimental values.
\cite{das18} attempted to improve on this using different sizes of $\rm{H_2O}$ clusters containing up to six~$\rm{H_2O}$ molecules to
calculate {the} BEs of $100$~species, including radicals of astrochemical interest, and noted that with increasing cluster size, {the} calculated BE values start to converge toward the experimentally obtained values.
\cite{shim18} introduced molecular dynamic (MD) simulations to describe the ASW cluster of 20~$\rm{H_2O}$ molecules and their interactions with atomic C, N, and O using quantum chemistry.
\cite{ferr20} employed density functional theory (DFT) using {B3LYP-D3(BJ)} (for closed-shell molecules) and M06-2X (for radical species) functionals and the Ahlrichs' triple-zeta quality VTZ basis set, supplemented with a
double set of polarization functions (A-VTZ*), to calculate {the} BEs of $17$~molecules and $4$~radicals (consisting of up to $8$~binding sites each), taking both crystalline and amorphous {periodic} ice models (consisting of $60$~$\rm{H_2O}$) into consideration.
They find good agreement with the previously mentioned studies.
In a similar vein, \cite{perr22} and \cite{mart24} investigated the BEs of various S- and N-bearing molecules and radicals adsorbed on similar water ice models using similar DFT and higher-level composite methodological approaches.
\cite{tina22} developed a computational framework to simulate the formation of interstellar ice mantles through H$_2$O accretion. They used the ONIOM (Our own N-layered Integrated molecular Orbital and molecular
Mechanics) approach to compute the BE distribution of NH$_3$ for an ASW ice model composed of 200~H$_2$O molecules. Their results highlight the complexity of adsorption sites and the potential for multiple BE regimes.
In a later study, \cite{tina23} calculated the distribution of the H$_2$O BE on the same ASW grain model and integrated it into a protoplanetary disk model. They assessed its effect on the position of the water snow line and estimated the water content of potential planetesimals at different distances from the disk center.
\cite{bari24} expanded on this by investigating the distribution of the H$_2$S BE on the same ASW ice surfaces, emphasizing the importance of considering a multitude of binding sites.
\cite{dufl21} applied the ONIOM high-quantum-mechanic (QM) part and a low-QM part hybrid method to calculate the BEs for a series of atoms, small molecules, and radicals on both crystalline and amorphous ice represented by a cluster of about $150$~$\rm{H_2O}$ molecules obtained via classical MD and electronic structure methods. \cite{same21} used the ONIOM QM/MM (MM denotes molecular mechanics) method to calculate the BE for ten binding sites of the $\rm{CH_3O}$ radical on both crystalline and amorphous ice clusters containing $162$~$\rm{H_2O}$ molecules ($49$ in the QM and $113$ in the MM region).
\cite{enri22} computed {the} BEs of radicals using DFT calculations on two ice water cluster models made of $33$ and $18$~$\rm{H_2O}$ molecules.
\cite{bovo22} computed $21$~BE distributions
of interstellar molecules and radicals on an amorphized set of $15-18$~water clusters of $22$~molecules each, for a total of $225-250$~unique binding sites.
\cite{piac22} published {the} BEs of ten P-bearing species on $1-3$~$\rm{H_2O}$ molecule clusters using DFT (M06-2X/aug-cc-pVDZ level of theory).
\cite{hend23} reported computational searches for optimal structures, benchmark binding, and condensation energies for sets of neutral, radical, cationic, and anionic molecules of astrochemical relevance with clusters of N~$=1-4$~$\rm{H_2O}$ molecules.
They demonstrated excellent agreement (with discrepancies of less than $\sim 200$~K) between the DFT ($\omega$B97X-V/def2-svpd and $\omega$B97X-V/def2-qzvppd levels
of theory) and coupled-cluster (CCSD(T)/aug-cc-pVTZ level of theory) BEs, which validates the DFT protocols. This could make the DFT the primary method for studying larger clusters.

The aforementioned studies predominately focused on the closed-shell species, where all the electrons are paired. In contrast, open-shell species (such as radicals) pose technical challenges due to convergence issues in self-consistent field (SCF) procedures.
The energy surface of radicals is more complex, with multiple local minima, making it harder to find the global minimum; additionally, the unpaired electrons result in a high electron correlation effect, requiring a more sophisticated treatment.
Experimentally determining the desorption data of unstable radicals is more challenging due to difficulties in their preparation and deposition, and their tendency to react before desorbing.
The chemical energy released upon a reaction can also trigger a desorption event \citep{mini22}. Therefore, focusing on stable species is more common due to the difficulties in handling molecular radicals experimentally.

The present work provides {the} BEs on various amorphous water ices
for some open-shell neutral radicals that can be directly implemented into astrochemical models.
In the past, there was a prevailing belief that radicals only diffused at relatively high temperatures \citep[>30 K;][]{garr08,das15a}. However, contemporary research suggests that complex organic molecules can be generated through non-diffusive chemistry \citep{schi19,jin20,iopp21}.
Since water is the major constituent of a grain mantle in molecular clouds \citep{kean01,das10,das11,das16}, we performed BE computations of the radicals CH,
NH, OH, SH, CN, NS, and NO, using nine different ASW clusters constructed with 20 $\rm{H_2O}$ molecules ($\rm{[H_2O]_{20}}$).
We also made a detailed comparison of the calculated BEs and previously available experimental or theoretically obtained BE values.

This paper is organized as follows. 
We discuss computational details and methodology in Sect.~\ref{sec:comp}.
In Sect.~\ref{sec:result_disc} results are reported and discussed in detail.
{Astrophysical implications are described in Sect.~\ref{sec:implication},} and we conclude in Sect.~\ref{sec:con}.


\section{Computational details and methodology} \label{sec:comp}
The BE is usually seen as a local property arising from the electronic interaction between the substrate (grain surface or adsorbent) and the species 
deposited on it (adsorbate).
For a bounded adsorbate, the BE is a positive quantity. It is defined as the difference in electronic
optimized energy between the separated radical ($E_{rad}$) plus the substrate ($E_{sub}$) and the adsorbed radical on the substrate
($E_{complex}$):
\begin{equation} \label{eq:BE}
    BE = E_{sub} + E_{rad} - E_{complex}.
\end{equation}
For our BE investigation, we considered nine different structures {of} $\rm{[H_2O]_{20}}$ ASW clusters (see Table~A.1, available on Zenodo) as the adsorbents.
{Each ASW ice model was created by cutting out a block from a large unit cell of crystalline ice. They were heated to 300~K using MD annealing calculations based on classical force fields. To prevent water molecules from escaping the cluster, 20 water molecules were selected using the TIP3P model, and droplet simulations were run with a spherical restriction of 20 $\AA$. Following 100~ps simulations at 300~K, the nine distinct structures were rapidly cooled to a low temperature of 10~K \citep{shim18}.}
These ASW structures are a good compromise between accuracy and computational time and/or cost for several radicals. Also, ice in astrophysical environments, such as icy satellites, comets, planetary rings, and interstellar grains, is mainly amorphous, and the main component of the ice mantle is $\rm{H_2O}$ \citep{boog15}.
Moreover, some recent studies \citep{germ22,tina22} suggest that a large
number of $\rm{H_2O}$ molecules in the ASW cluster  (at least a few hundred) could mimic the interstellar ice mantle. 
However, these types of ASW are too large to handle the complete quantum chemical treatment (geometric optimization and harmonic
frequency evaluation).
For a large number of binding sites, the $\rm{[H_2O]_{20}}$ cluster would be a reasonable choice for studying the BE.
The smaller size of ASW clusters results in the formation of \enquote{under-coordinated} water molecules at their surface, making them more likely to have adsorbates and higher BE values.
{The larger ASW clusters, on the contrary, tend to have the maximum number of internal H bonds. This gives them a more spherical shape, and thus they have fewer available strong adsorption sites for the adsorbates \citep{mart24} due to a reduction in edges and/or cavities in the model.}

All the quantum chemical calculations were performed using the \texttt{Gaussian~09} and \texttt{Gaussian~16} suite of programs \citep{fris09,fris13,fris16,fris19}.
The nine different structures of $\rm{[H_2O]_{20}}$ clusters we examined were fully optimized based on DFT calculations using the $\omega$B97X-D functional approach \citep{chai08}, in which the van der Waals  interaction (dispersion correction) is empirically incorporated.
{The choice of $\omega$B97X-D was motivated by its demonstrated accuracy in describing systems with unpaired electrons, such as radicals. The range-separated hybrid nature of this functional enables a balanced treatment of both electrostatic and dispersion interactions, which are crucial for accurately modeling radical-water complexes. While benchmarking against higher levels of theory is beyond the scope of this work, it is important and we intend to do so in future studies.}
A valence triple-$\zeta$ plus polarization and diffuse functions, 
namely the 6-311+G(d,p) basis set, was adopted for the geometry optimization and the BE evaluation.
We verified that a fully optimized electronic ground-state structure (i.e., the local {minimum}) is a stationary point (with a nonnegative
frequency) via harmonic vibrational frequency analysis. All the BEs were evaluated with the harmonic vibrational zero-point energy corrections to study the molecular vibration at low temperatures.
Unless otherwise stated, all the BEs discussed in this work are the zero-point-energy-corrected BEs.
We corrected for the basis set superposition error (BSSE) using the counterpoise correction (CPC) method \citep{boys70} to get an idea of the BSSE dependence of the system (see Sect.~\ref{sec:bsse}).

We used $\rm{[H_2O]_{20}}$ {clusters} to simulate the interstellar ASW surface and investigate the influence of H-bond cooperativity on radical BEs as well as the effect of {the} BE on binding site variation \citep{bovo20,ferr20,dufl21,bovo22}.
The accessible radical binding sites were chosen based on the available dangling-H ($d$-H) positions on ASW $\rm{[H_2O]_{20}}$ cluster surfaces.
The BE is controlled by electrostatic attractions, orbital interactions, and Pauli repulsion. As a result, each radical species can have a range of possible BEs \citep{mini22}.
As per the information provided in Table~A.1 (available on Zenodo), a total of 75 $d$-H sites are available; these sites are distributed among ASW $\rm{[H_2O]_{20}}$ clusters 1 to 9 and have 7, 7, 9, 10, 8, 9, 8, 9, and 8 $d$-H sites, respectively.
Using a weak H-like bond, geometry optimizations of the complex structures were performed after manually connecting the radical with the $d$-H.
During this weak bond connection, we always took the largest differences in electronegativity values (see \autoref{tab:en}) between two binding elements of the atom in the radical and $d$-H atom in the $\rm{[H_2O]_{20}}$ cluster.
{The electronegativity of each atom contributes to the formation of molecular dipoles, which can be strategically utilized to construct the initial adsorption geometries manually. By following the principle of electrostatic complementarity, we always connected the $d$-H atom of the $\rm{[H_2O]_{20}}$ cluster with the atom in the radical that has the higher electronegativity. This approach leverages the natural tendency of atoms with higher electronegativity to attract electrons or electron density toward themselves, resulting in stronger dipole interactions and more stable adsorption configurations.}
The electronic ground-state spin multiplicity of each of the seven radicals was checked and is noted in \autoref{tab:BE}. All the considered radicals are {doublets} (with one unpaired electron) except NH, which is in the triplet state (with two unpaired electrons of the same spin).
{The coordinates of all the representative structures are publicly available in a GitHub repository\footnote{\url{https://github.com/milansil/ASW_radicals_data.git}}.}

\begin{table}
\caption{{Pauling electronegativity values for the elements.}
\label{tab:en}}
\centering
\begin{tabular}{cc}
    \hline
    \hline
{\bf Species} & {\bf Electronegativity} \\
    \hline
 H  & 2.1 \\
 C  & 2.5  \\
{S} & 2.5  \\
 N  & 3.0  \\
 O  & 3.5  \\
\hline
\end{tabular}
\end{table}

It is to be noted that the optimized energy and the calculated BEs obtained from the \texttt{Gaussian} program are not always accurate, due to stationary point convergence in frequency domain calculations.
This is because the root mean square (RMS) and the maximum force are calculated analytically in the SCF and frequency domain. However, SCF numerically solves the maximum and RMS displacement; on the contrary, the post-SCF (frequency) solves both of them analytically. Hence, some deviation of $\pm 1$~K can be observed. Due to time constraints and high computational costs, the values were taken even when only the maximum and RMS forces were converged in the frequency domain, and the frequencies are all positive.


\section{Computational results and discussion} \label{sec:result_disc}

\subsection{Binding energy}
\begin{figure}
    \centering
    \includegraphics[width= 0.55\textwidth]{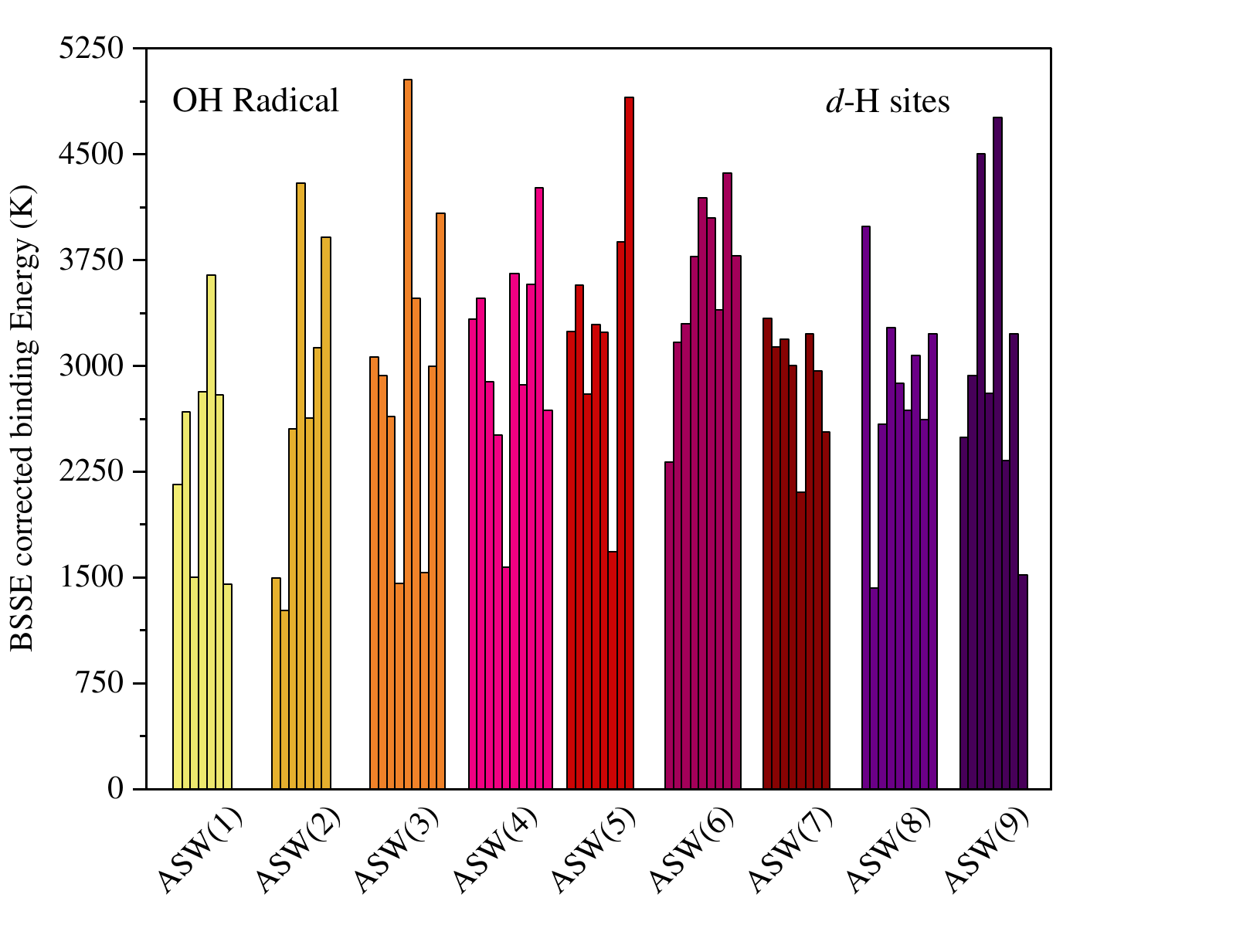}
    \caption{BSSE-corrected BE values of all $75$ $d$-H positions for the nine different ASW $\rm{[H_2O]_{20}}$ clusters. The corresponding values are presented in Table~A.2 (available on Zenodo).}
    \label{fig:d-H}
\end{figure}

\begin{figure}
    \centering
    \includegraphics[width=0.5\textwidth]{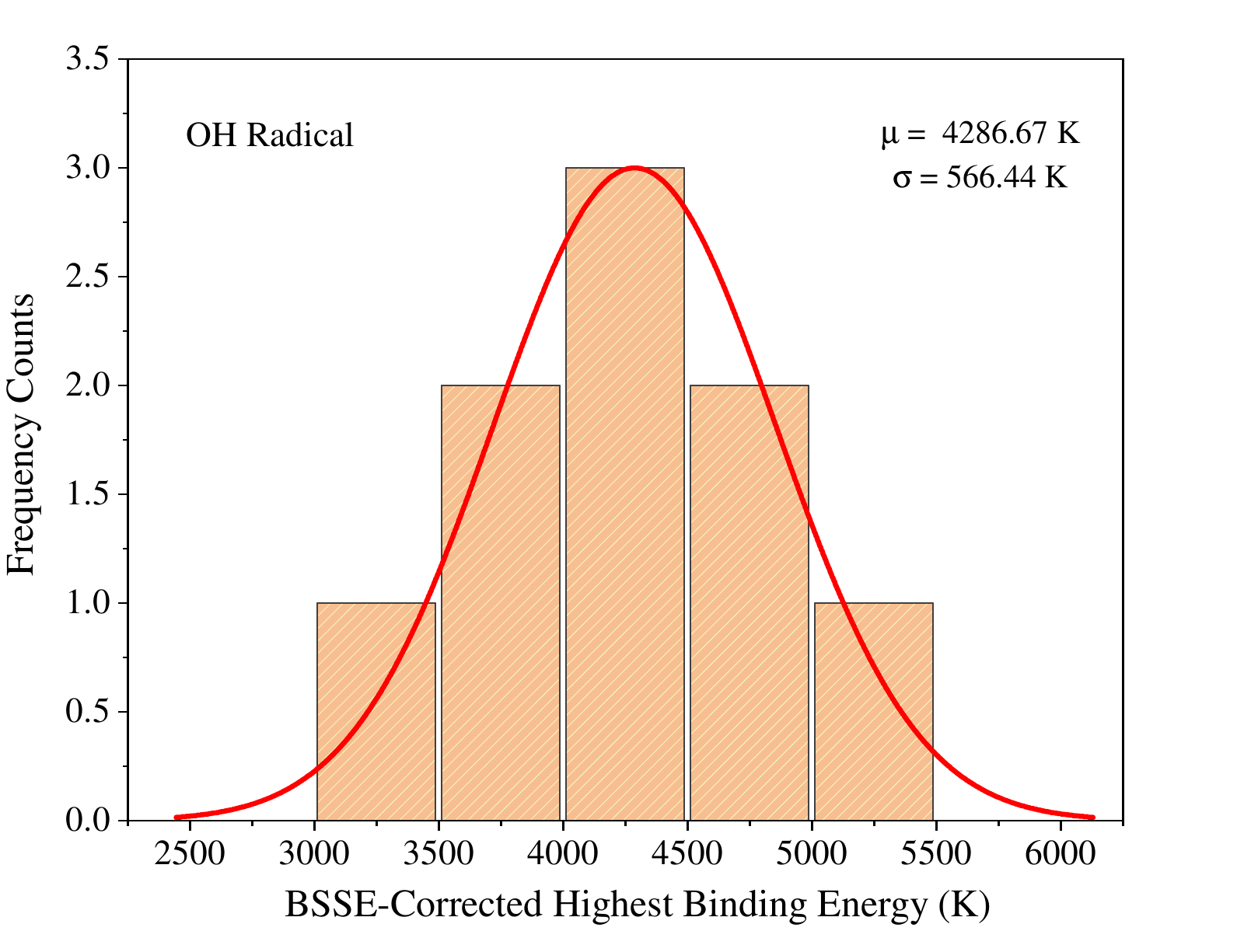}
    \caption{Gaussian distribution for the highest BE values of the OH radicals over the highest $d$-H binding sites for the nine {ASW $\rm{[H_2O]_{20}}$ clusters.}} \label{fig:bsse_bd}
\end{figure}

First, we calculated the optimization energy of nine {ASW} $\rm{[H_2O]_{20}}$ clusters (presented in Table~A.1, available on Zenodo), along with their optimized structures. The {ASW(1)} cluster is, on average, $\sim 2958$~K more stable than the other eight {ASW clusters}.
We then calculated the BE of the OH radical at the $75$ $d$-H binding sites of nine ASW clusters.
All the calculated {(+) BSSE BE} values are plotted in Fig.~\ref{fig:d-H} and tabulated in Table~A.2 (available on Zenodo).
Instead of taking the average BE over all $75$ binding sites, 
{we considered the average over the highest $d$-H binding sites in each cluster}. This averaging would be more appropriate for estimating the BE at a low temperature (interstellar condition).
We obtain average BE values of $4287$~K and $4741$~K for the OH radical with and without considering BSSE corrections, respectively. The locations of the BEs depend on the binding sites and are broadly distributed (see Table~A.2, available on Zenodo).
{To verify the distribution of the BSSE-corrected BE values over the highest $d$-H binding sites in each cluster},
we organized them in a distribution plot with a bin width of 500~K (see Fig.~\ref{fig:bsse_bd}). The distribution is {well reproduced by an overlay of the Gaussian probability distribution function:}
\begin{equation}
    \mathcal{G}(x,\mu,\sigma) = \frac{1}{\sqrt{2\pi\sigma^2}}\exp \bigg(-\frac{(x-\mu)^2}{2\sigma^2}\bigg)\,,
\end{equation}
where $\mu$ is the mean and $\sigma$ is the standard deviation.
The fitted parameters for the BE value distribution are the mean ($\mu$) $\approx 4287$~K and standard deviation ($\sigma$) $\approx 566$~K. This implies that the Boltzmann distribution (uniform distribution to the highest BE site at $0$~K) holds for each local region, which is further supported by the fitting using the Boltzmann model equation (see Fig.~A.1, available on Zenodo).
On the contrary, the BE distribution considering all 75 sites (BSSE-corrected and uncorrected) {does not} mimic a Boltzmann distribution (non-Gaussian type), as depicted in Fig.~A.2 (available on Zenodo).

To minimize the computational burden, we performed the calculation of the BE for the remaining six radicals (i.e., CH, NH, SH, CN, NS, and NO) by considering only the {ASW(1)} $\rm{[H_2O]_{20}}$ cluster. This particular {ASW cluster is the most stable (lowest energy) of the nine ASW $\rm{[H_2O]_{20}}$ clusters} (see Table~A.1, available on Zenodo). Then, we selected the highest physisorbed BE values among the seven binding sites available in {ASW(1)} (see Table~A.3, available on Zenodo).
It is interesting to note that all the radicals except NO (i.e., CH, NH, OH, SH, CN, and NS) give the highest BEs (considering both physisorption and chemisorption) with the fifth $d$-H position of {the ASW(1)} cluster.
Since we only considered the ASW(1) cluster to estimate the BE of five of these six radicals (excluding OH), we wanted to examine the effect on the average BE when considering all nine {ASW clusters}. For OH, we have both the average BE of the nine {ASW clusters} and the maximum BE from {the ASW(1) cluster}. We calculate a ratio of 1.177, which indicates a $\approx15\%$ deviation; we got this from the OH radical calculation, by dividing the BSSE-corrected average BE (4287~K) by {the highest BSSE-corrected BE (3643~K) of the ASW(1) cluster} (see Table~A.2, available on Zenodo). Consequently, we applied this scaling factor to the other six radicals to determine the effect on the average BE over all nine ASW clusters, as we only estimated {the} BEs for these species using {the} ASW(1) cluster.
As mentioned earlier, the standard deviation ($\approx$ 566 K) for OH was calculated using the Gaussian distribution over the nine highest BEs obtained for the nine ASW clusters, and thus 566~K can be treated as an error bar for {the} BE of OH. To get an idea of this error bar for the other six radicals, a standard deviation ($\approx$ 786 K) for OH was calculated by taking the seven binding sites of the ASW(1) cluster.
Finally, a scaling factor of $0.721$ was obtained by taking their ratio. 
For the other six radicals, we obtained the standard deviation like we did for OH radical, using the seven binding sites of the ASW(1) cluster and then multiplying them by the scaling factor of $0.721$ to get an estimate of the standard deviation for the nine ASW clusters (see Fig.~A.5, available on Zenodo).
{It should be noted that this calculation does not take the significantly higher BE values for CH and CN into account, as they are considered special cases.}
The final BSSE-corrected scaled BE values, along with the error bars, are noted in the last column of Table \ref{tab:BE}.

\begin{table*}
\caption{BE calculations for radicals and their comparison.} \label{tab:BE}
\resizebox{\linewidth}{!}{\begin{tabular}{|c|c|c|c|c|c|c|c|c|}
\hline
{\bf Sl.} & {\bf Radicals} & {\bf Formula} & {\bf Ground-state} & \multicolumn{5}{c|}{\bf BE in Kelvin} \\
\cline{5-9}
{\bf No.} &&&{\bf spin multiplicity} & {\bf PENTEADO$^a$} & {\bf DAS$^b$} &{\bf KIDA$^*$ database} & {\bf Experimental} & {\bf This work} \\
\hline
\hline
1. & Methylidyne radical & CH & Doublet & $590\pm295$ & --- & $925^{c,g}$ & --- & {$2044 \pm 94$} \\
2. & Imidogen & NH & Triplet & $542\pm270$ & 1947 & $2378^{c,g}, \ 2600\pm780^d$ & --- & {$2234 \pm 178$} \\
3. & Hydroxyl radical & OH & Doublet & $3210\pm1550$ & 3183 & $2850^{c,g},\ 4600\pm1380^{d},\ 5698^e$ & $4600^f,\ 1656-4760^h$ & {$4287\pm 566$} \\
4. & Mercapto radical & SH / HS & Doublet & $1350\pm500$ & 2221 & $1450^{c,g},\ 2700\pm810^{d}$ & --- & {$2482 \pm 314$} \\
5. & Cyano radical & CN & Doublet & $1355\pm500$ & 1736 & $1600^{c,g},\ 2800\pm840^{d}$ & --- & {$1278 \pm 41$} \\
6. & Nitrogen monosulfide & NS & Doublet & $1800\pm500$ & 2774 & $1900^{c,g}$ & --- & {$2619 \pm 198$} \\
7. & Nitric oxide & NO & Doublet & $1085\pm500$ & $886-1988$ & $1600^{c,g},\ 1600\pm480^{d}$ & --- & {$704 \pm 94$} \\
\hline
\end{tabular}}
\tablefoot{$^a$ \cite{pent17},
$^b$ \cite{das18},
$^c$ OSU gas-phase database,
$^d$ \cite{wake17},
$^e$ \cite{mini22},
$^f$ \cite{duli13},
$^g$ \cite{garr13},
$^h$ \cite{he14}, 
$^*$ KIDA: \url{https://kida.astrochem-tools.org/}.}
\end{table*}

\subsection{BSSE correction}
\label{sec:bsse}
The BSSE is intuitively a highly plausible concept.
The BSSE correction using the CPC method accounts for the superposition error caused by the sloppy treatment of the basis set for each monomer of a big cluster, as the intermolecular distance varies.
Even if this inconsistency could be eradicated entirely, errors related to the incompleteness of the basis set, known as the basis set completeness error (BSCE), would remain. Hypothetically, the BSSE and the BSCE would be zero in the complete basis set limit. The BSSE and BSCE often have opposite signs \citep{dunning2000road}.
For small and medium basis sets around the van der Waals minimum, the (negative) BSCE cancels (part of) the (positive) BSSE \citep{sheng11}.
Ignoring one basis set fault while accounting for the other can produce imbalanced results.
According to \cite{sheng11}, diffuse basis functions can help reduce the BSSE. However, they also pointed out that diffuse basis functions can reduce the BSCE. Therefore, the partial cancelation of these errors still applies, and the extensive use of diffuse basis functions {does not} lead to a different conclusion.
\cite{jensen} recently pointed out that the most significant source of error for smaller basis sets is {the BSCE}. Furthermore, he noted that these errors are dominant in dispersion-type interactions and that including CPC can, in such cases, lead to a deterioration of the results.

The recent studies by \cite{perr22}, \cite{bovo22}, and \cite{tina23} suggest that BE values can be erroneous for bigger systems without BSSE correction. Considering these contradictory statements, and to reduce the computational cost and make the calculations possible, we used only one small basis set, 6-311+G(d,p), and thus cannot extrapolate toward the complete basis set limit. Hence, we relied on an estimate to determine the influence of the BSSE on the highest physisorption BEs of the radicals.

\begin{table}
\caption{BSSE-corrected ($+$) and uncorrected ($-$) BEs for different radicals for the ASW(1) $\rm{[H_2O]_{20}}$ cluster.
\label{tab:BSSE}}
\centering
\begin{tabular}{c c c c c}
    \hline
    \hline
{\bf Sl}  & {\bf Radicals} & \multicolumn{3}{c}{\bf $\omega$B97X-D / 6-311+G(d,p)} \\
{\bf No.} &  & {\bf $-$BSSE} & {\bf $+$BSSE} & {$\rm{\frac{\bf -BSSE}{\bf +BSSE}}$} \vspace{1mm}\\
    \hline
1. & CH  & 1892 & 1737 & 1.0892 \\
2. & NH  & 2083 & 1898 & 1.0975 \\
3. & OH  & 3990 & 3643 & 1.0952 \\
4. & SH / HS & 2416 & 2109 & 1.1456 \\
5. & CN & 1192 & 1086 & 1.0976 \\
6. & NS & 2606 & 2225 & 1.1712 \\
7. & NO & 931 & 598 & 1.5569 \\
\hline
\end{tabular}
\end{table}

\begin{figure}
    \centering
    \includegraphics[width=0.5\textwidth]{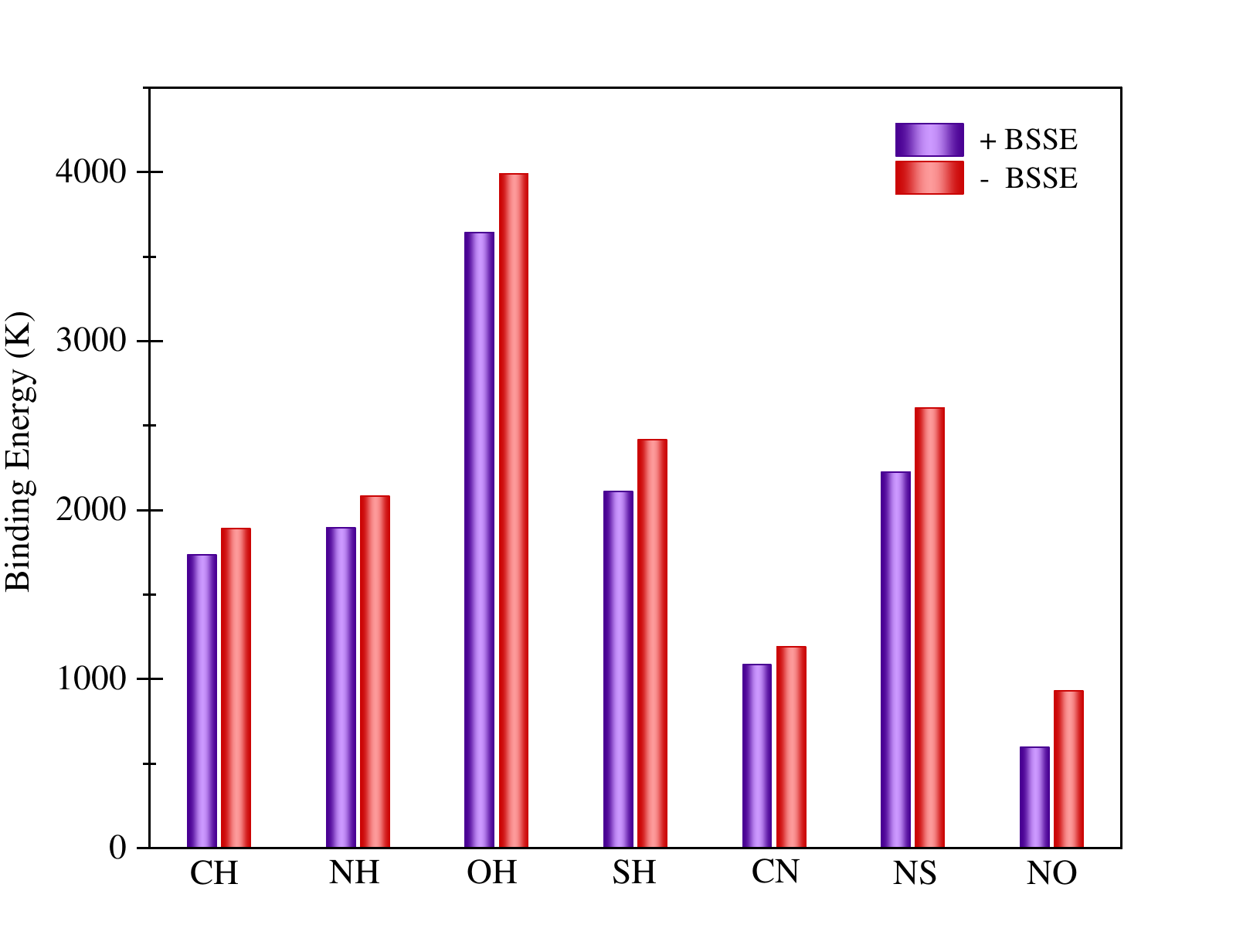}
    \includegraphics[width=0.5\textwidth]{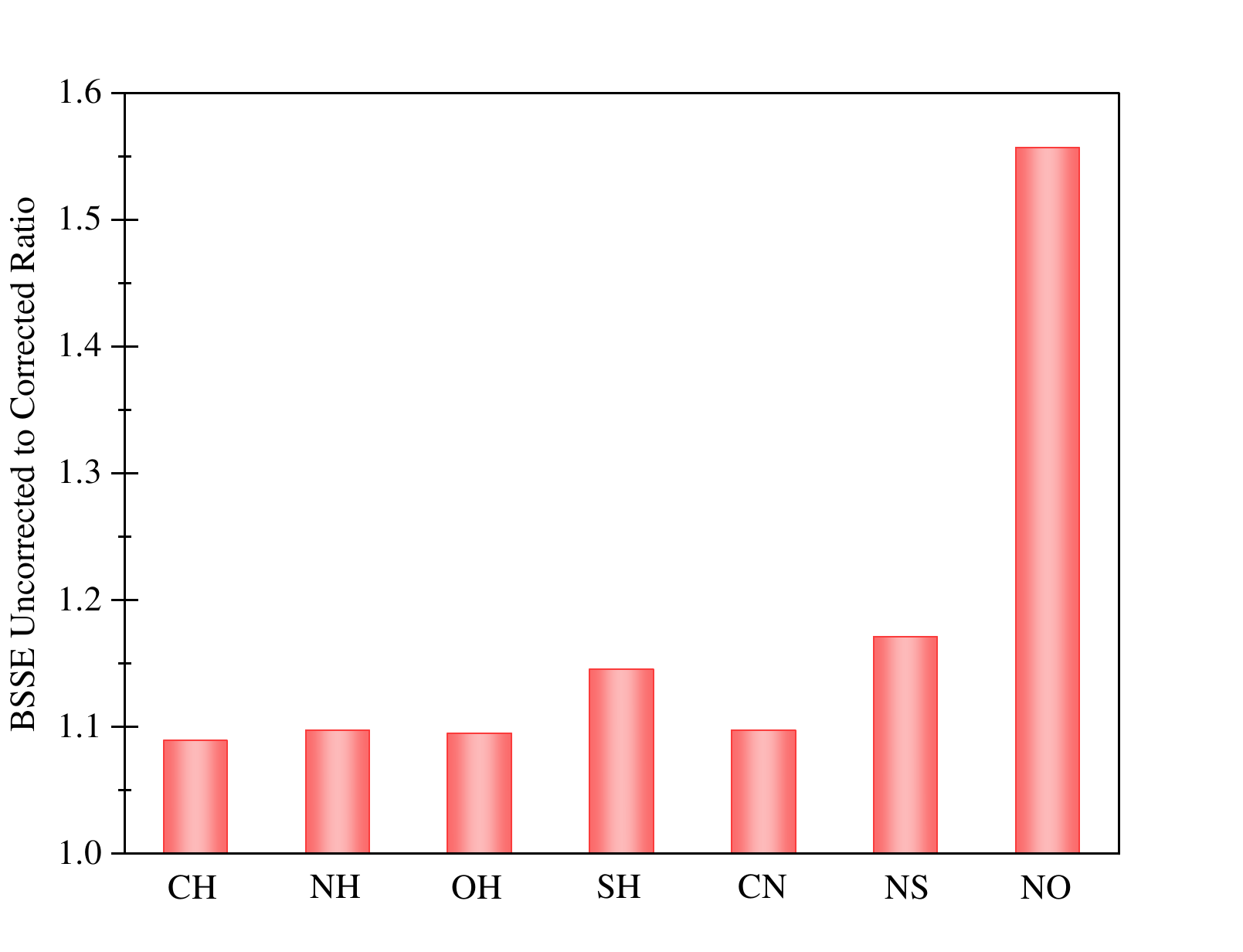}
    \caption{Bar plots for BSSE-corrected ($+$) and uncorrected ($-$) highest BE values (upper panel) and the ratio between the uncorrected and corrected BE values (lower panel) for seven radicals.}
    \label{fig:radical_BSSE}
\end{figure}

In \autoref{tab:BSSE} we compare the BE results obtained by carrying out BSSE correction using the CPC method and the BSSE-uncorrected BE for the $d$-H binding sites with the highest BE found for all seven radicals with {the ASW(1)} cluster.
It is to be noted that for the CH radical, the highest BEs are due to the chemisorption {process} (see Table~A.3, available on Zenodo).
Chemisorption refers to the formation of a chemical bond between two atoms of the radical and water cluster.
The structural geometries (see Fig.~A.3, available on Zenodo) suggest that for the CH radical, the cluster geometries get broken, and consequently, CH$_2$OH and CHOH+H formation happens at sites 5 and 7 of the {ASW(1)} cluster, respectively.
This is similar to the case found for C and CH in \cite{wake17}, \cite{shim18}, and \cite{molp21}.
We considered the highest BE value among only the physisorbed energy values for the sake of calculation consistency.
The result is less evident for the CN radical (an exceptionally large BSSE-corrected BE of 6317~K at site 5 in the ASW(1) cluster) as no structural change can be seen. 
It is thought that CN forms a nonclassical 2c-3e bond (a hemibond) between the unpaired electron on the CN carbon atom and the lone pair of electrons of the oxygen atom of the water surface instead of forming H bonds with the surface {\citep{rimo18,mart24}} and is most probably the reason for the {significantly large BE value.}
For the case of CN, we considered 1086~K as the highest BSSE-corrected  BE value for consistency.
To better understand the dependence of the BSSE on the BE, the BSSE-corrected and uncorrected BEs are plotted side by side in Fig.~\ref{fig:radical_BSSE} and compared with the ratio by which the BSSE-corrected BEs are lower than the uncorrected one for all seven radicals.
{From \autoref{tab:BSSE}, we see that the average ratio is $\approx1.179$, considering the $d$-H binding sites with the highest BE found for all seven radicals with the ASW(1) cluster.} A quantitative study on the BSSE dependence of the OH radical taking all the binding sites in the ASW(1) cluster into consideration was also performed (see Table~A.4, available on Zenodo). The BSSE-corrected and uncorrected values and the uncorrected-to-corrected ratio are plotted in Fig.~A.4 (available on Zenodo).
The average uncorrected-to-corrected ratio for OH is $\approx 1.119$. The BE values with BSSE adjustments are consistently lower than those without BSSE corrections for all cases.
Unless otherwise stated, all BEs discussed hereafter are BSSE-corrected BEs.

\subsection{Comparison of {the} BEs with previous studies}

\begin{figure}
    \centering
    \includegraphics[width=0.5\textwidth]{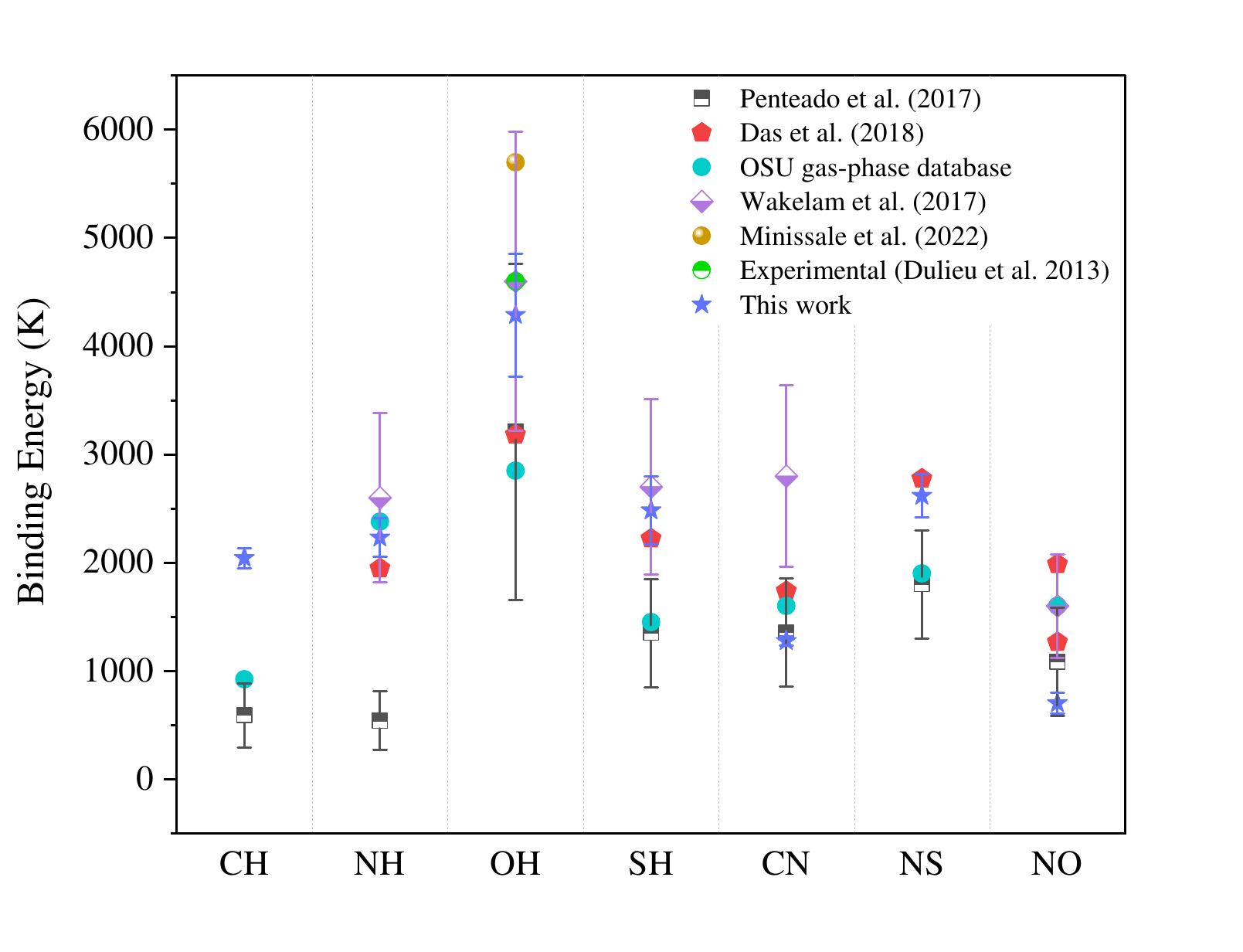}
    \caption{Comparison of {the} BEs computed in the present work (blue stars) and those based on the available literature and frequently used databases noted in \autoref{tab:BE}.
    The solid lines go from the minimum to the maximum BE values for each species. Only the OH radical has reported experimental data (green circle). The dashed gray vertical lines guide the eye to identify the corresponding species.
    \label{fig:all_BE}}
\end{figure}

The experimental investigation of radical binding properties is complex since, during the heating ramp of conventional temperature programmed
desorption, the radical has an enormous chance to diffuse and react, making the experimental evaluation of the desorption parameters almost impossible.
The obtained BE values depend strongly on the chemical composition and morphology of the substrate, as well as the monolayer or multilayer regime used in the experiment \citep{he16}.
To the best of our knowledge, no experimental studies have provided the BEs of the radicals considered in this study, except for a handful of experimental determinations for {the} BE of OH on amorphous silicate surfaces.
{Therefore}, we do not have many constraints to benchmark our computed values.
Moreover, comparing the theoretical BEs with experimentally determined ones is not straightforward.
We must, therefore, rely on existing theoretical  reference values.
In \autoref{tab:BE} we compare our calculated BE with other available experimental, computational, and public database values.
To better visualize the comparison, we plot all the tabulated data on a common diagram in Fig.~\ref{fig:all_BE}.
To use a coherent set of homogeneous data for comparison, we chose the BE values recommended by \cite[except for CH]{das18}  and \cite{pent17} and used the KInetic Database for Astrochemistry (KIDA) for many of the radicals.
\cite{pent17} present a list of {the} recommended BEs, which includes data from previous studies and an uncertainty range for each value.
The uncertainty assigned is set to half the BE for species with less than 1000~K (CH and NH) and 500~K for all other cases (SH, CN, NS, and NO).
For all the radicals considered here except OH, the provided average BEs are based on the work of \cite{hase93} and \cite{aika96}. \\

{OH}: The binding properties of the OH radical are difficult to study experimentally due to its tendency to diffuse and react, leading to the formation of ASW on the bare surface during conventional temperature programmed
desorption heating ramps. Only a couple of existing experimental studies have reported the BE for hydroxyl radical on bare amorphous silicates: 4600~K \citep{duli13} and $1656-4760$~K \citep{he14}. \cite{pent17} later recommended {BE =} $3210\pm1550$~K, based on the distribution determined by \cite{he14}, taking the average with uncertainty to cover this range.
However, as far as we know, no experimental study considering ASW surfaces has been published to date.
Several other existing reference values are available for OH.
The Ohio State University (OSU) gas-phase database gives 2850~K, assuming half the value of $\rm{H_2O}$ \citep{garr06}.
A systematic empirical approach by \cite{wake17} provided $4600\pm1380$~K using the $\rm{[H_2O]_1}$ ASW model.
\cite{das18}, using quantum calculations and the $\rm{[H_2O]_4}$ ASW model, find 3183~K.
\cite{ferr20} determined the range $1816-6230$~K using the amorphous 2D slab ice model, generated from the $\rm{[H_2O]_{20}}$ {structures} from \cite{shim18}.
There are some estimated average values ranging between 4300~K \citep[considering both crystalline and ASW substrates;][]{miya20} and 4990~K \citep[between 2321~K and 7770~K on top of a crystalline water ice;][]{same17}.
\cite{dufl21} find an average best estimate of $5199 \pm 1938$~K between minimum 2936~K and maximum 8924~K for the amorphous ice model.
\cite{enri22} calculated 2911~K and 5376~K for the $\rm{[H_2O]_{18}}$ and $\rm{[H_2O]_{33}}$ ASW ice models.
Finally, after examining the available studies, \cite{mini22} recommended a value of 5698~K.
Very recently, \cite{hend23} computed {the} BEs of OH to the $\rm{H_2O}$ monomer, dimer, trimer, and tetramer as 1922~K, 3900~K, 4252~K, and 3674~K, respectively, using the CCSD(T)/aug-cc-pVTZ level of theory and 1907~K, 3905~K, 4433~K, and 3840~K, respectively, using  the {DFT-$\omega$B97X-V/def2-qzvppd} level of theory.
They noticed that the H-bond distances between the O atom of OH and the H atom of $\rm{H_2O}$ shorten notably when an $\rm{H_2O}$ molecule is added to the cluster.
As shown in Fig.~\ref{fig:all_BE}, our calculated average BE of $4287\pm566$~K is well within the range of existing experimental, theoretical, and recommended values, and is close to the central value.\\

{CH}: As per our knowledge, no computational or experimental BE value exists for the methylidyne (also called carbyne) radical to date.
\cite{wake17} attempted to consider CH in their work but noticed the formation of strong (partly) covalent bonding with a single H$_2$O molecule. However, they did not provide any chemisorption value. We calculated its BE considering seven~$d$-H sites of the ASW(1) $\rm{[H_2O]_{20}}$ cluster (see Table~A.3, available on Zenodo).
We notice that two~$d$-H sites favor chemisorption (see Fig.~A.3, available on Zenodo), whereas the other five~sites favor physisorption. {The \ce{C-H} chemical bond lengths of $1.372~\AA$ and $1.513~\AA$ (which are longer than the typical covalent bond length of $1.08~\AA$) are formed at these two chemisorption binding sites.}
A recent study by \cite{Nguyen21} on the reaction rate of CH with H$_2$O vapor suggests that in the reaction pathway, the first step is the barrier-less association
leading to a pre-reactive complex, which has a BE of {$8.93$~kcal~mol$^{-1}$} ($\sim 4494$~K). This pre-reactive complex  (HC$^{\delta+}$–${}^{\delta-}$OH$_2$) is formed via
a polar interaction between a lone pair orbital of the O atom and an empty p-orbital of the C atom. Subsequently, the pre-reactive complex  can isomerize to CH$_2$OH if CH is inserted into an O–H bond of H$_2$O.
A similar chemisorption behavior of a C atom with H$_2$O was recognized by previous studies \citep{wake17,shim18,molp21}.
The large dipole moment \citep[1.46~D;][]{phel66} could be the reason for the strong long-range interaction, which would mean it is reactive to neutral $\rm{H_2O}$ molecules.
As seen from \autoref{tab:BE} and Fig.~\ref{fig:all_BE}, our proposed average BE of $2044\pm94$~K is higher than the estimated range of $590\pm295$~K (based on educated guesses) provided in \cite{pent17}, as well as the 925~K provided by the OSU gas-phase database.
To clarify, our average BE was calculated without taking  the chemisorbed BE values into account (see Table~A.3, available on Zenodo). \\

{NH}: We report an average BE for imidogen, or nitrene, of {$2234\pm178$~K}, which is in good agreement with the OSU database value of 2378~K and within the range $2600\pm780$~K proposed by \cite{wake17}.
Our calculated value is well above the recommended value of $542\pm270$~K \citep{pent17} and higher than the value of 1947~K computed by \cite{das18}.
\cite{mart20} report BEs of 4222~K for a crystalline water-ice surface.
Our value agrees well with the best-estimated average value of $2437 \pm 1126$~K for the amorphous ice model \citep{dufl21}.
\cite{enri22} recently calculated {the} BEs of NH as 1560~K and 3909~K on the $\rm{[H_2O]_{18}}$ and $\rm{[H_2O]_{33}}$ ASW ice models, respectively, considering a single binding site; this is a simplistic assumption given that different surface binding sites are available and, accordingly, a distribution of BEs exists.
{\cite{enri22} also noted (from the private communication with B. Martinez-Bachs) the BE range $1320 - 5410$~K for ASW ice surfaces, centered at around 2410~K.
However,} the very recent calculations by \cite{mart24} show a range of BEs for the ASW model, varying from a minimum of 1167~K to a maximum of 4692~K, with an average of 2464~K, which is also in close agreement with our value.\\

{SH/HS}: We report the average BE for mercapto, or the sulfanyl radical, as {$2482\pm314$~K}, which is within the range $2600\pm780$~K proposed by \cite{wake17}.
However, the other values -- 1450~K (OSU database), $1350\pm500$~K \citep{pent17}, 2221~K \citep{das18}, and $2195\pm931$~K \citep{perr22} -- are somewhat lower than our computed average value.     \\

{CN}: Our computed average BE value for the cyano radical ($1278\pm41$~K) {is} consistent with the existing values of 1600~K (OSU database), $1355\pm500$~K \citep{pent17}, and 1736~K \citep{das18}, but not with the larger value of $2800\pm840$~K proposed by \cite{wake17}.
However, an exceptionally high BE of 6317~K is found for site 5 in the {ASW(1) $ \rm{[H_2O]_{20}}$ cluster}. This value was not included in the average calculation. {This high BE value coincides with the range of BEs proposed by \cite{mart24}: 2605~K to 7792~K, with an average of 4856~K. The hemibond nature is most probably the reason for this inconsistency.} \\

{NS}: Unlike the case of the CH radical, our calculated average BE for the nitrogen monosulfide radical, $2619\pm198$~K, is higher than the available estimated range of $1800\pm500$~K \citep{pent17}, as well as the 1900~K provided by the OSU gas-phase database.
The value of 2774~K calculated by \cite{das18} is somewhat close to our calculated average value.
\cite{perr22} calculated {an} average BE of 2026~K, a minimum of 1234~K, and a maximum of 2955~K for {an} amorphous ice model. \\

{NO}: Our computed average BE value of $704\pm94$~K for the nitric oxide radical is well within the range $1085\pm500$~K proposed by \cite{pent17}  as well as the range calculated by \cite{das18} for a different ASW ice model ($568-1988$~K). The KIDA database suggests a higher value of $1600\pm480$~K as proposed by \cite{wake17}.
Very recently, \cite{hend23} computed {the} BEs of the NO to $\rm{H_2O}$ monomer, dimer, trimer, and tetramer as 367~K, 1147~K, 438~K, and 1354~K, respectively, using the CCSD(T)/aug-cc-pVTZ level of theory and 252~K, 1022~K, 649~K, and 1087~K using {the DFT-$\omega$B97X-V/def2-qzvppd} level of theory. Similar to NH, the results of \cite{mart24} give an energy range of 291~K to 965~K, with an average of 712~K, which strongly agrees with our results. The low BE of NO on the surface is primarily governed by dispersive interactions. This is due to the almost negligible dipole moment of NO, which limits the contribution of electrostatic interactions to the overall binding. \\

It is important to note that the small fluctuations observed in the calculated BEs compared to literature values can be attributed to several factors, including the specific computational method employed, the adopted ice model, and the particular binding sites investigated in this study.
Open-shell species pose significant computational challenges due to multi-reference characteristics and the need to account for static and dynamic correlation effects. Accurately capturing the true energy of these species requires a sensitive and careful approach, often involving multi-reference wave function theories (e.g., the coupled cluster method) and perturbation methods (e.g., the M\o{}ller-Plesset method). Although these techniques can provide highly accurate results, they come with a substantial computational cost.
The combination of the $\omega$B97X-D functional and the 6-311+G(d,p) basis set offers a practical balance between accuracy and efficiency. The $\omega$B97X-D functional is crucial for handling the complex electronic structures inherent to radicals and is a versatile tool for studying a wide range of systems.
The 6-311+G(d,p) basis set, with its valence triple-$\zeta$ quality and added polarization functions, further enhances the accuracy by providing a flexible and detailed description of electron density, especially where electron distribution is uneven due to unpaired electrons. This combination is not necessarily the best or most accurate option available. Still, it is a well-considered choice that delivers a high degree of accuracy while keeping computational costs relatively low.
Furthermore, we believe that our methodology, with careful structural optimization and analysis, is robust enough to be applied to other radical species and, more generally, to species with challenging electronic structures.

\section{Astrophysical implications} \label{sec:implication}
It is crucial to comprehend the impact of the newly computed BEs on interstellar ice chemistry, as some of them differ from the values used in the past few decades. We outline the implications of these computed values below.

\subsection{Chemisorption}
Some of our binding sites exhibit chemisorption for CH (a chemical bond is formed with one oxygen atom of a water molecule of the cluster). Figure~A.3 (available on Zenodo) depicts the formation of CH$_2$OH and CHOH$_2$ by chemisorption. There is a possibility that the chemisorbed CH$_2$OH could produce CH$_3$OH via the Eley-Rideal mechanism by capturing an H atom from the gas phase. Alternatively, a chemisorbed H atom could react with CH$_2$OH to form methanol.
Recent experiments by \cite{grie24} suggest the formation of H$_2$O at a high temperature (approximately 85 K) by chemisorbed H and O. Additionally, \cite{grie23} reported the formation of H$_2$ up to a temperature as high as 250 K, suggesting the feasibility of hydrogenation reactions at higher temperatures than previously thought. 
{Contrary to previous understandings of methanol formation through the surface hydrogenation of CO, \cite{molp21} and \cite{tsug23} suggest that before CO freeze-out, atomic carbon deposited on ASW could form H$_2$CO.} \cite{tsug23} pointed out that at temperatures below 20~K, some of the accreted carbon atoms are quickly converted to formaldehyde. The remaining physisorbed atoms are partially changed to chemisorbed and hydrogenated to CH$_4$. Physisorbed carbon atoms can diffuse at temperatures between 20~K and 30~K, promoting insertion reactions with other carbon atoms or species.
Future observations with the \textit{James Webb} Space Telescope (JWST) could provide additional details about the significance of chemisorption in forming complex organic molecules in the ice phase.

\subsection{Implication of {the} computed BEs for the astrochemical model}
Recently, using Atacama Large Millimeter/submillimeter Array (ALMA) observations, \cite{cout19} detected multiple transitions of HONO in the low-mass protostellar system IRAS 16293–2422. This system consists of three protostars, two located toward source A and one toward source B \citep{maur20}. The identification by \cite{cout19} specifically focuses on the protostar associated with source B.
They also attempted to explain the abundances 
of HONO-related species and their ratios using a chemical model. 
However, they were unable to account for all the observed features. 
In their study, \cite{cout19} estimated abundance ratios of HONO/NH$_2$OH and HONO/HNO to be greater than or equal to $2.3$ and $3$, respectively. Using their chemical model, they successfully explain the abundance of HONO, N$_2$O, and NO$_2$. 
However, their model resulted in an overproduction of NH$_2$OH and HNO, leading to lower {HONO/NH$_2$OH} ($1.7 \times 10^{-3}$) and HONO/HNO (0.01) ratios compared to the observed values. Our study suggests a weak binding of NO with the water substrate. It would be interesting to see how the estimated BEs would affect the modeled abundances of these species.

We utilized the Rokko code \cite{furu15,furu17} for our chemical modeling, which encompasses three phases: gas, icy grain surface, and homogeneous bulk ice mantle. Our active ice layers were limited to the top four monolayers. All layers, including the active ice layers, were considered part of the bulk ice mantle. The gas- and ice-phase networks are based on the work of \cite{garr13}. The active layers undergo photodissociation and the photo-desorption of species. A constant reactive desorption of all the species by a factor of 0.01 is assumed, resulting in a single product.
MD studies have determined that approximately 10\% of the formation energy is retained after the formation of H$_2$S from HS+H \citep[and references therein]{bari24}. However, based on experimentally determined reaction desorption probabilities for H$_2$S, HDS, and D$_2$S \citep{oba19,furu22}, we used a reactive desorption factor of 0.03. We also used a standard cosmic-ray ionization rate of $1.3 \times 10^{-17}$~s$^{-1}$.

\begin{figure}
    \centering
    \includegraphics[width=0.5\textwidth]{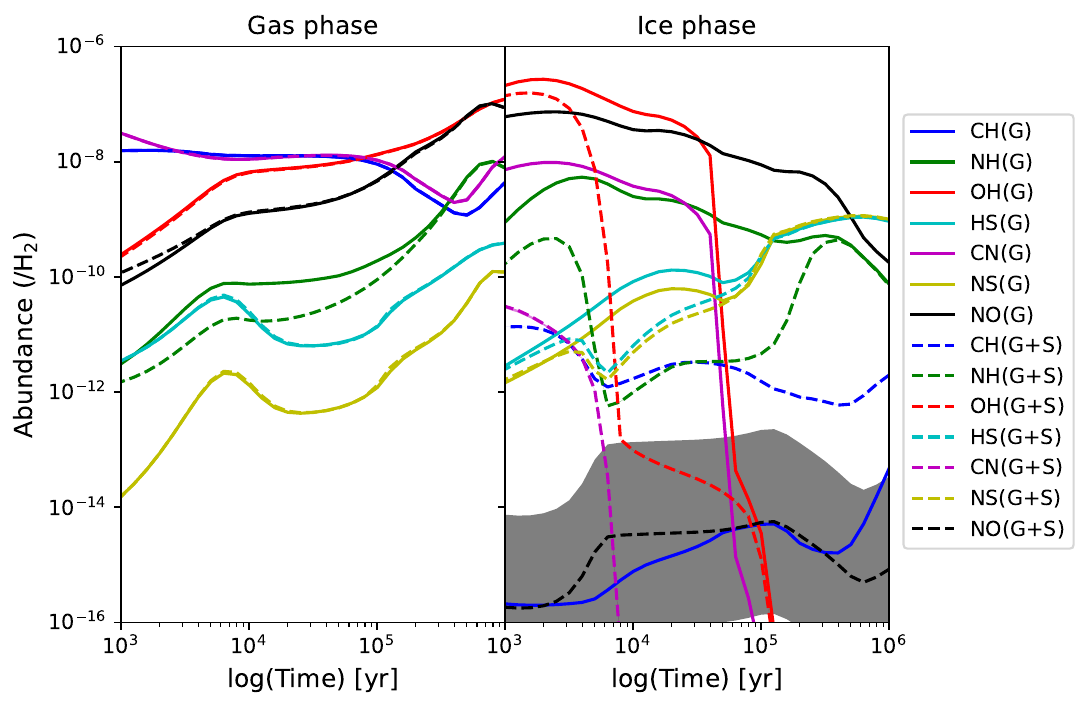}
      \includegraphics[width=0.5\textwidth]{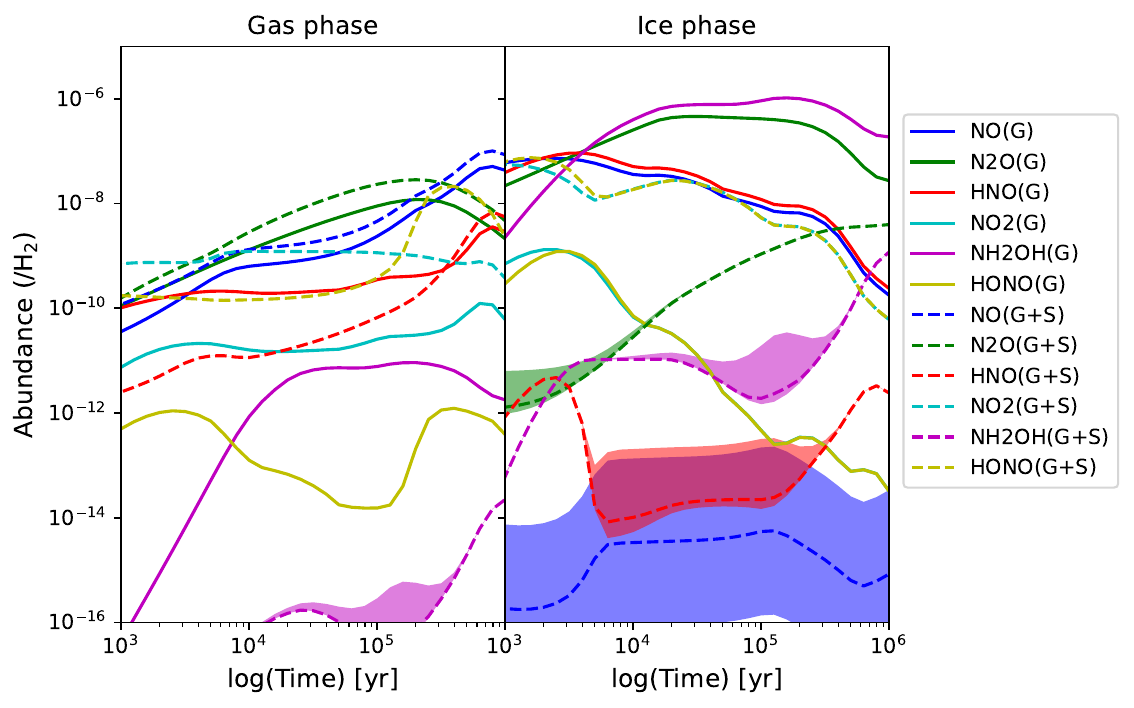}
    \caption{Changes in the abundances of gas- and ice-phase radicals over time in a dark cloud. The solid lines represent the results obtained using the BE values from \citet[``G'']{garr13}, while the dashed curves represent the results obtained using our calculated BE values (``G+S''). The filled curve around the calculated BE curves shows the abundances when the errors on the calculated BEs are taken into account, as presented in Table \ref{tab:BE}.}
  
    \label{fig:dc}
\end{figure}

\subsubsection{Dark cloud \label{sec:dc}}
For the dark cloud model, we started with a hydrogen number density, $n_H=2 \times 10^4$ cm$^{-3}$, a visual extinction ($A_V$) of 10 mag, and a gas and ice temperature of 10 K. 
The initial elemental abundances of species were taken from \cite{garr17}.
The upper panel of Fig.~\ref{fig:dc} shows the time evolution of the radicals. Here, ``G'' indicates that all the BE sets are from \cite{garr13};  ``G+S'' indicates that the BEs for the radicals are taken from this work and the other BEs from \cite{garr13}. In this work we used a default diffusion-to-BE ratio of 0.4.
Since no estimated BEs were available for HONO, H$_2$NO$_2$, and H$_3$NO$_2$, we used our estimated values of 2918 K, 4205 K, and 3044 K, respectively, in both ``G'' and ``G+S.''
Changes in {the} BE would have a minor effect on the gas-phase abundances of these radicals. However, they would have a major effect on the ice-phase abundances of these radicals, especially for the ice-phase NO and CH. The NO ice abundance is reduced by a factor of almost $10^5$ at $10^6$~years. Meanwhile, the ice-phase abundance of CH increases by a factor of 100.
The reduction in the abundance of NO is due to the weak binding of NO obtained in this work ($\sim 704$ K) compared to the earlier estimate \citep[1600~K;][]{wake17}. The increase in the surface abundance of CH radicals is due to the higher estimated  BE ($\sim 2044$ K) compared to previous work \citep[925 K]{garr13}.
The shaded lines on the graph indicate the uncertainty of our BE values from Table \ref{tab:BE}. Since NO has a very weak binding with the water substrate, its low uncertainty (about 94 K) in {the} BE estimation could cause a significant change in its ice-phase abundance (see the black shaded area in the top right panel of Fig.~\ref{fig:dc}).

The lower panel of Fig. \ref{fig:dc} presents the time evolution of the NO-related species.
It clearly shows that with the present BE estimation of NO, we can expect a favorable formation of gas-phase HONO in a dark cloud. On the contrary, the formation of gas-phase NH$_2$OH is heavily affected, which could be useful in explaining the observed HONO/NH$_2$OH ratio in star-forming regions \citep{cout19} discussed in Sect.~\ref{sec:hc}. The shaded regions are the results when the uncertainty in {the} BE calculations {is} taken into consideration. The obtained abundances and abundance ratio with HONO are noted in Table \ref{tab:abun}.
Due to the weak binding of NO obtained in this study, we find a higher abundance of the ice-phase NO$_2$ through the reaction $\rm{NO-gr + O-gr \rightarrow NO_2-gr}$, which results in a favorable formation of HONO-gr by ${\rm H-gr+NO_2-gr \rightarrow HONO-gr}$.
The formation of NH$_2$OH-gr occurs through successive hydrogenation reactions: ${\rm NO-gr \xrightarrow{H-gr} HNO-gr \xrightarrow{H-gr} HNOH-gr \xrightarrow{H-gr} NH_2OH-gr}$. When our BE value of NO is used, due to the high abundance of atomic oxygen on the grain compared to the atomic hydrogen, NO-gr is mainly channelized to form $\rm{NO_2-gr}$, resulting in a decrease in the formation of $\rm{NH_2OH-gr}$. On the contrary, it facilitates the HONO-gr formation. The lower panel of Fig.~\ref{fig:dc} presents the observed significant uncertainties in the ice-phase abundances of HNO and NH$_2$OH due to the uncertainty surrounding the BE of NO ($\sim$ 94~K).

\begin{figure}
    \centering
    \includegraphics[width=0.5\textwidth]{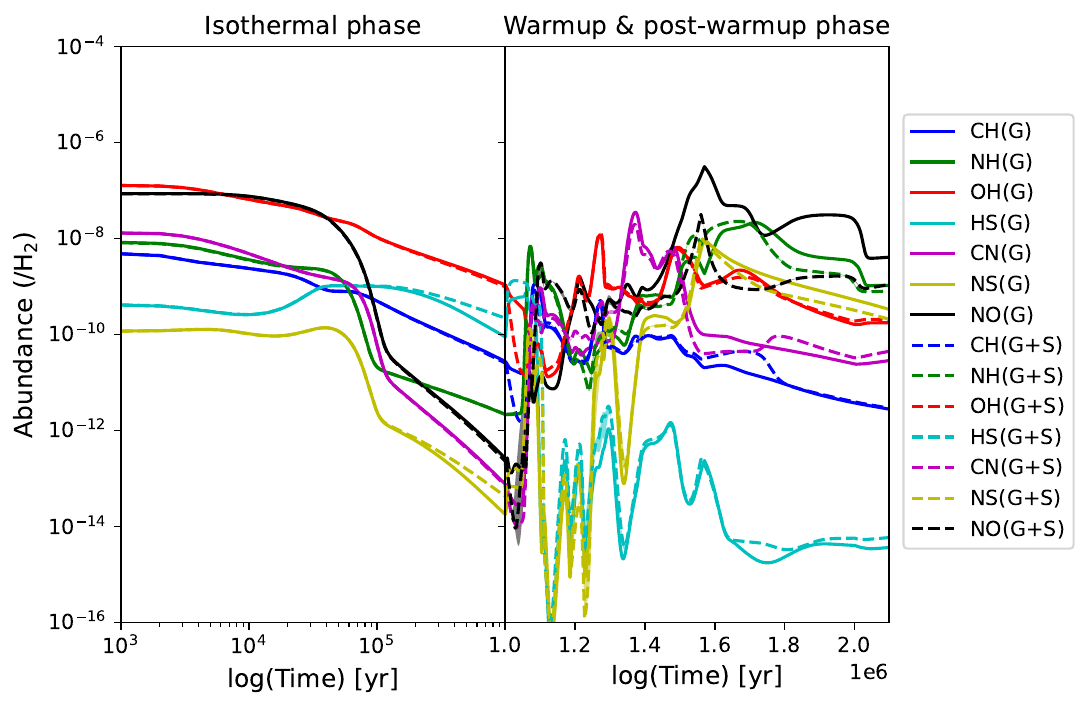}
      \includegraphics[width=0.5\textwidth]{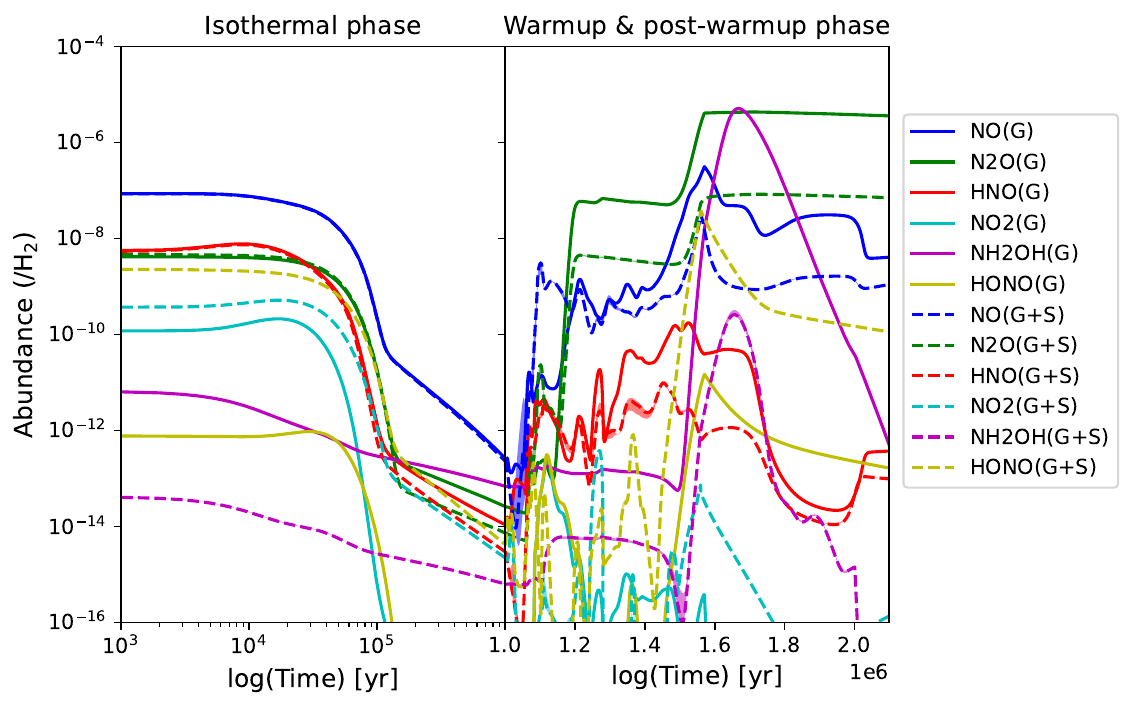}
    \caption{Time evolution of the abundances of gas-phase NO-related species in the hot core. Solid lines represent the BE used from \citet[``G'']{garr13}, and dashed curves when our calculated BEs are used (``G+S'').}
    \label{fig:hc}
\end{figure}

\begin{table*}
\centering
\caption{{Peak abundances and abundance ratios for NO-related species obtained from our model and a comparison of peak abundance ratios with the available observation.}
\label{tab:abun}}
{\begin{tabular}{c c c c c}
    \hline
    \hline
{\bf Species}  & {\bf Dark cloud} & {\bf Hot core } &  {\bf Observation} \\
\hline
HONO& $7.7 \times 10^{-13g}/2.3 \times 10^{-9s}$ & $1.5 \times 10^{-11g}/3.8 \times 10^{-8s}$   & --- \\
NO& $8.6 \times 10^{-8g}/8.6 \times 10^{-8s}$ & $3.2 \times 10^{-7g}/3.2 \times 10^{-8s}$   &  ---  \\
N$_2$O& $4.2 \times 10^{-9g}/4.7 \times 10^{-9s}$   &$4.4 \times 10^{-6g}/8.3 \times 10^{-8s}$    & --- \\
HNO& $5.8 \times 10^{-9g}/5.6 \times 10^{-9s}$ &   $1.5 \times 10^{-10g}/7.1 \times 10^{-12s}$   & --- \\
NO$_2$&  $1.2 \times 10^{-10g}/3.8 \times 10^{-10s}$  &  $2.5 \times 10^{-13g}/2.3 \times 10^{-13s}$  & --- \\
NH$_2$OH&   $3.6 \times 10^{-12g}/2.2 \times 10^{-14s}$ & $5.0 \times 10^{-6g}/2.1 \times 10^{-10s}$   & --- \\
CH$_3$OH& $9.2 \times 10^{-9g}/9.2 \times 10^{-9s}$   & $3.1 \times 10^{-5g}/3.3 \times 10^{-5s}$   & --- \\
HONO/NO& $8.9 \times 10^{-6g}/0.03^s$   & $4.7 \times 10^{-5g}/1.2^s$ &  $4.5 \times 10^{-2c}$\\
HONO/N$_2$O& $0.0002^g/0.5^s$   & $3.4 \times 10^{-6g}/0.4^s$ &  $\le 2.3 \times 10^{-2c}$\\
HONO/HNO& $0.0001^g/0.4^s$ &   $0.08^g/3781^s$ & $\ge 3^c$ \\
HONO/NO$_2$&  $0.006^g/6.1^s$  & $56^g/9.8 \times 10^{4s}$  &  $\ge 4.5 \times 10^{-2c}$\\
HONO/NH$_2$OH&  $0.22^g/1.0 \times 10^{5s}$  & $2.9 \times 10^{-6g}/143^s$ &  $\ge 2.3^c$\\
HONO/CH$_3$OH&  $8.4 \times 10^{-5g}/0.25^s$  & $4.8 \times 10^{-7g}/0.001^s$ & $9 \times 10^{-5c}$ \\
\hline
\end{tabular}}
\tablefoot{$^c$\citet{cout19} in the low-mass protostellar source IRAS 16293–2422~B. \\
$^g$ When {the BEs} from \citet{garr13} {are} used. \\
$^s$ When {the BEs} calculated here {are} used (otherwise those from \citealt{garr13} are used).}
\end{table*}

\subsubsection{Hot corino \label{sec:hc}}

 We further examined a low-mass star-forming region to assess the impact of our proposed BEs on abundance estimations. We considered three stages: the isothermal collapsing stage, the warm-up stage, and the post-warm-up stage. During the isothermal collapsing stage, the cloud was allowed to collapse from a minimum density of $n_{min}=3 \times 10^3$~cm$^{-3}$ to a maximum density of $n_{max}=10^7$~cm$^{-3}$ in 1 million years, with a constant gas temperature of 10~K and an ice temperature of 10~K. The initial abundance for this model was taken from the final abundance of the dark cloud model discussed in Sect.~\ref{sec:dc}. The visual extinction in the first stage was allowed to vary with the density following the relation provided by \cite{garr11}, resulting in a variation of $A_V=2-446$. In the warm-up stage, the density and visual extinction parameters were consistently set to their maximum values for an additional 1 million years, and the temperature gradually rose to a peak of $\sim 200$~K. During the post-warm-up stage, the parameters listed above remained at their maximum values for an additional $10^5$~years.

The upper panel of Fig.~\ref{fig:hc} shows the time evolution of the abundances of gas-phase radicals. With the new BE value, the gas-phase abundance of NO (dashed black curve) decreases significantly during the warm-up stage.
In the lower panel of Fig.~\ref{fig:hc}, we show the abundances of NO and its related species separately. The uncertainties in the BE calculations are also considered here as discussed in the context of Fig.~\ref{fig:dc}.
In \autoref{tab:abun}, we compare the peak abundances (peak value is taken beyond the isothermal stage) and the peak abundance ratio of some NO-related species obtained from our simulations with that observed by \cite{cout19}.
With the new BEs (represented by the dashed curves), we observe a significant reduction in the abundances of NO, N$_2$O, HNO, and NH$_2$OH. Conversely, there is a surge in the abundance of HONO. 

NH$_2$OH is primarily produced in the cold ice phase through the successive sequential hydrogenation of NO-gr, as discussed in Sect. \ref{sec:dc}, and it can sublimate to the gas phase after around $1.7 \times 10^6$~years, depending on the BE \citep[6806 K;][]{garr13}. Due to the weak binding of NO, its abundance in the ice phase decreases as it forms NO$_2$, leading to a decrease in NH$_2$OH. We estimate a BE of 2918~K for HONO, so it is expected to be released earlier than NH$_2$OH in the gas phase. However, an additional radical-radical pathway ($\rm{OH-gr+NO-gr\rightarrow HONO-gr}$) is responsible for ice-phase HONO production in warmer regions. The low BE of NO obtained here further facilitates this radical-radical ice-phase reaction.

In our study, with our estimated BE for NO, we obtain a lower abundance of NH$_2$OH and HNO (see Fig.~\ref{fig:hc} and \autoref{tab:abun}). 
We obtain peak abundance ratios of HONO/NH$_2$OH and HONO/HNO of 143 and 3781, respectively, consistent with the observation.
\autoref{tab:abun} shows that the other ratios from the observations seem to be between that found using the previous model \citep{cout19} and this model. This may indicate that the diffusion activation energy of NO was overestimated in the previous one and underestimated in the present.

\section{Conclusions} \label{sec:con}
This study aimed to calculate the BE of certain interstellar radicals on cluster surfaces of ASW. We employed a fully quantum chemical approach to examine a large number of binding sites for all the considered radicals using {a suitable} ASW cluster model. The significant findings of our work are as follows:

\begin{enumerate}

    \item We estimated the BEs of seven diatomic radicals using a pure quantum chemical approach with high-level water clusters and methods, and we discuss the implications of the computed BEs for astrochemical models.

    \item We carefully and extensively employed the BSSE correction using the CPC method. This correction is crucial, especially with small and medium-sized basis sets, for obtaining accurate and reliable estimates of the BEs between radical species and the ASW cluster.
    
    \item 
    Our results suggest that, most of the time, we have physisorption, indicating weak bonding between radicals and the water molecule.
    However, in addition to physisorption, we find chemisorption of CH with the $\rm{[H_2O]_{20}}$ cluster in two cases. In these cases, a chemical bond is formed between CH and an oxygen atom of a water molecule. Figure~A.3 (available on Zenodo) illustrates the production of CH$_2$OH through the chemisorption of CH on a water substrate. This warrants the potential formation of methanol through hydrogenation in warmer regions. Future observations with the JWST could address this issue. Moreover, our estimated value of the BE (physisorption) of CH is 2-3 times higher than the available estimate. Hence, we suggest using these values for astrochemical models.

    \item {We obtain a significantly lower BE value of NO (704~K), which is in excellent agreement with a very recent study \citep[average value of 712~K with the ASW model;][]{mart24} compared to that used earlier (1600~K).}
    Our obtained BEs of NO can help explain the observed HONO/NH$_2$OH and HONO/HNO ratio obtained in the B component of the low-mass protostellar binary IRAS 16293–2422.

\end{enumerate}

\section*{Data availability}
The data underlying this article (in the appendix) are made available under a Creative Commons Attribution license on Zenodo: doi:\href{https://doi.org/10.5281/zenodo.13388691}{10.5281/zenodo.13388691}.

\begin{acknowledgements}
    The authors thank the referee for the valuable insights and constructive comments that significantly improved the manuscript.
    M.S. acknowledges financial support through the European Research Council (consolidated grant COLLEXISM, grant agreement ID: 811363).
    N.I. acknowledges the support of FONDECYT Regular Grant 1241193 and Vicerector\'{i}a de Investigaci\'{o}n y Postgrado (VRIP). A.D. acknowledges the support of MPE for sponsoring a scientific visit to MPE.
\end{acknowledgements}

\bibliographystyle{aa}
\bibliography{references}

\end{document}